\documentclass[prl,twocolumn,showpacs]{revtex4}
\usepackage[dvips]{graphicx}
\usepackage{latexsym}
\usepackage{amsmath}
\usepackage{amsfonts}
\usepackage{amssymb}
\usepackage{bm}
\usepackage{color}
\usepackage{mathrsfs}
\begin{document}
\newcommand{\fig}[2]{\includegraphics[width=#1]{#2}}
\newcommand{{\vhf}}{\chi^\text{v}_f}
\newcommand{{\vhd}}{\chi^\text{v}_d}
\newcommand{{\vpd}}{\Delta^\text{v}_d}
\newcommand{{\ved}}{\epsilon^\text{v}_d}
\newcommand{{\vved}}{\varepsilon^\text{v}_d}
\newcommand{{\bk}}{{\bf k}}
\newcommand{{\bq}}{{\bf q}}
\newcommand{{\tr}}{{\rm tr}}
\newcommand{\pprl}{Phys. Rev. Lett. \ }
\newcommand{\pprb}{Phys. Rev. {B}}

\title{Antiferromagnetic Chern insulators in non-centrosymmetric systems}

\author{Kun Jiang,$^{1,2,3}$  Sen Zhou,$^{2}$  Xi Dai,$^{3}$ and Ziqiang Wang$^{1,3}$}
\affiliation{$^1$ Department of Physics, Boston College, Chestnut Hill, MA 02467, USA}
\affiliation{$^2$ CAS Key Laboratory of  Theoretical Physics, Institute of Theoretical Physics, Chinese Academy of Sciences, Beijing 100190, China}
\affiliation{$^3$ Beijing National Laboratory for Condensed Matter Physics and Institute of Physics,
	Chinese Academy of Sciences, Beijing 100190, China}
\date{\today}

\begin{abstract}

We investigate a new class of topological antiferromagnetic (AF) Chern insulators driven by electronic interactions in two-dimensional systems without inversion symmetry. Despite the absence of a net magnetization, AF Chern insulators (AFCI) possess a nonzero Chern number $C$ and exhibit the quantum anomalous Hall effect (QAHE).
Their existence is guaranteed by the bifurcation of the boundary line of Weyl points between a quantum spin Hall insulator and a topologically trivial phase with the emergence of AF long-range order. As a concrete example, we study the phase structure of the honeycomb lattice Kane-Mele model as a function of the
inversion-breaking ionic potential and the Hubbard interaction. We find an easy $z$-axis $C=1$ AFCI phase and a spin-flop transition to a topologically trivial $xy$-plane collinear antiferromagnet.
We propose experimental realizations of the AFCI and QAHE in correlated electron materials and cold atom systems.

\typeout{polish abstract}
\end{abstract}


\maketitle

Electronic band insulators can be characterized by their spin-dependent band topology and symmetry protected gapless edge states \cite{kane,qi}. Without time-reversal symmetry, the band topology is characterized by an integer Chern number $C$ in two-dimensions (2D) \cite{kane,qi,haldane}. Such Chern insulators include the quantum Hall and quantum anomalous Hall effect (QAHE) insulators, supporting $C$ number of chiral edge modes and quantized Hall conductance $\sigma_{xy}=C e^2/h$. In the presence of time reversal symmetry and spin-orbit interaction (SOI), the band topology changes to the one specified by a $Z_2$ number that produces 2D quantum spin Hall (QSH) and 3D topological insulators \cite{kane1,kane2,bhz}. In QSH insulators, $C=0$ but the spin Chern numbers $C_s=\pm1$.
There are two counter-propagating edge states with opposite spin-polarizations related by time-reversal symmetry.
Certain crystalline symmetry can also protect the band topology and edge states, leading to topological crystalline insulators \cite{fu}. These topological states are stable against weak electron-electron interactions and have all been observed experimentally recently.

In this paper, we study the topological properties of interaction-driven quantum states, focusing on {\em antiferromagnetic (AF) insulators} that are common in systems with strong local correlation. To the extent that the low energy physics of such AF insulators are adiabatically connected to band insulators in the magnetic unit cell and protected by the magnetic gap, we argue that they can indeed be topological insulators. In addition to the previously proposed $Z_2$ or AF topological crystalline insulators \cite{fang,liucx}, we show that there exists a class of AF Chern insulators (AFCI) exhibiting the QAHE despite having a filling fraction enforced zero total magnetization. We first provide a general discussion of the physical origin of AFCI due to {\em spontaneous} time-reversal symmetry breaking. Then we study a concrete example of the Kane-Mele model for the QSH insulator \cite{kane1,kane2} and include the electronic Hubbard interaction $U$. We show this Kane-Mele Hubbard model exhibits a $C=1$ AFCI phase for sufficiently large $U$. The resulting phases and phase transitions are studied.
Possible experimental realizations of the AFCI will be discussed.

An ideal AF insulator has zero total magnetization, i.e. ${\vec S}=\sum_{i\in{\cal M}}{\vec S}_i=0$, where ${\cal M}$ is the magnetic unit cell. It can be obtained by filling up an integer number of bands occupied by electrons of both spins in the magnetic Brouillon zone. For this to happen, the SOI must leave at least one spin-rotation invariant axis.
Denoting the latter as the spin $z$-axis, $S_z$ is a conserved quantum number. It implies that the expectation values $S_x=S_y=0$ and $S_z={1\over2}(N_\uparrow-N_\downarrow)$ where $N_{\uparrow,\downarrow}$ stand for the number of filled bands in each spin component. Hence, $S_z={1\over2}m$ where $m$ is an integer. Thus magnetic insulators are topologically protected by the band gap and filling fractions. Specifically, an AF insulator with $m=0$ occurs when half of the bands at half-filling, or one-quarter of the bands at quarter-filling, etc, are filled.
When spin-rotation symmetry is completely broken, e.g. by the presence of both SOI and Rashba coupling, perfect AF insulators with zero total spin are not protected; interactions in general produce ferromagnetic moments and the resulting insulators are thus ferrimagnets.

To understand the topological properties of AF insulators, it is useful to take a closer look at the interplay between time-reversal ${\cal T}$ and spatial symmetries $({\cal R})$ of the crystal.
Consider a two-sublattice antiferromagnet obtained by spontaneously breaking ${\cal T}$, spin-rotation, and translation by half a magnetic lattice vector ($t_{1\over2}$).
The loss of the nonunitary ${\cal T}$, which is crucial for defining topological insulators, can produce two types of topological AF insulators. In the first kind, the combined time-reversal and certain spatial operations ${\cal\widetilde T}\equiv{\cal T}\otimes{\cal R}$ remains a nonunitary symmetry with ${\cal\widetilde T}^2=-1$, capable of reinstating the magnetic counterpart of topological insulators. Indeed, the case of ${\cal\widetilde T}={\cal T}\otimes t_{1\over2}$
has been used to define $Z_2$ topological AF insulators, and ${\cal\widetilde T}={\cal T}\otimes (C_4,C_6)$ to AF topological crystalline insulators under the magnetic crystalline group \cite{fang,liucx}.
Such topological AF states can also exist in 2D; but they are of course Chern trivial. The second kind, to be studied in detail below, corresponds to 2D AFCI that arise when ${\cal T}$ is truly broken, i.e. when the time-reversed electronic states cannot be brought back to their original ones by any space-group operations. The latter usually requires noncollinear spin moments or the breaking of lattice inversion symmetry.
Despite being perfect antiferromagnets as defined above, they have a band topology with a nontrivial Chern number $C\neq0$ and exhibit the QAHE.

To illustrate the physical origin of AFCI, consider 2D non-centrosymmetric systems that break the inversion symmetry ${\cal I}$.
The operations of ${\cal I}$ and ${\cal T}$ on a quantum state $\vert k, \sigma\rangle$, where $k$ and $\sigma$ label the momentum and spin, are elementary:
${\cal I}|k,\sigma\rangle\rightarrow|\bar{k},\sigma\rangle$, $
{\cal T}|k,\sigma\rangle\rightarrow|\bar{k},\bar{\sigma}\rangle$, and
${\cal T}\otimes {\cal I}|k,\sigma\rangle\rightarrow|k,\bar{\sigma}\rangle$.
The Kramers doublet at a given $k$ requires ${\cal T}\otimes {\cal I}$ to be a symmetry. Breaking either ${\cal I}$ or ${\cal T}$ lifts the degeneracy and can produce the Weyl points in the dispersion as in 3D Weyl semimetals \cite{nielsen,wan,xu,weng}. Weyl points can also arise in 2D ${\cal T}$-invariant noncentrosymmetric systems. The simplest one corresponds
to two spin-valley locked Weyl points at $K(\downarrow)$ and $K'(\uparrow)$ described by the low energy Hamiltonian
\begin{eqnarray}
H_W&=&k_x\sigma_z\tau_x+k_y\tau_y,
\end{eqnarray}
with the momenta expanded around the valleys at $K^\prime=-K$. Here $\vec{\sigma}$ and ${\vec\tau}$ are the Pauli matrices with $\sigma_z$ denoting the valley/spin and $\tau_z$ the pseudospin states of the (sub)bands. Since the time-reversed spin states are locked to the valleys, $H_W$ is ${\cal T}$-invariant, but odd under ${\cal I}$. Unlike in 3D, such a 2D Weyl semimetal is critical and lives on the boundary between a topological and a trivial phase. 
For instance, if the same mass is introduced at each valley by adding a ${\cal T}$-invariant $m_I\tau_z$ to $H_W$, a $Z_2$ topological insulator with spin Chern number $C_s=1$ emerges for $m_I<0$, while a trivial band insulator (BI) obtains when $m_I>0$. This is the QSH realization
in the Kane-Mele model of the more general two-dimensional topological insulator (2DTI). On the other hand,
a mass $m_s\sigma_z\tau_z$ couples to each valley with opposite signs and breaks ${\cal T}$. Since the two valleys have opposite spin polarizations, one of them becomes topologically trivial while the other carries a nontrivial Chern number. Thus, a Chern insulator with $C=1$ is obtained. Remarkably, such a truly ${\cal T}$-breaking spin-mass can be generated spontaneously by a correlation-driven AF order,
resulting in the AFCI.
\begin{figure}
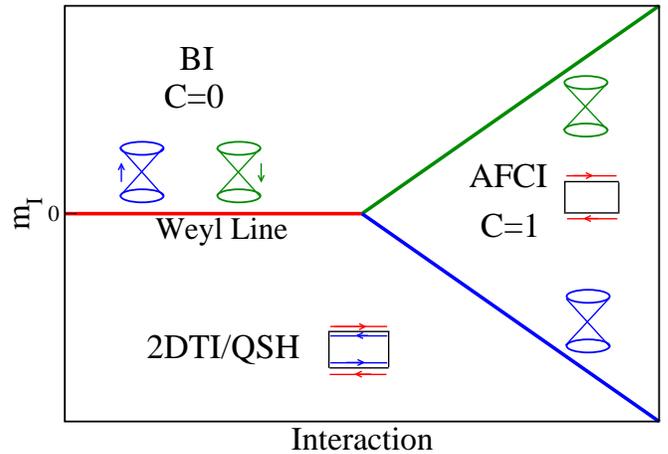

	\begin{center}
		\fig{3.4in}{weyl.eps}\caption{Schematic phase diagram near the line of Weyl points (red line).
Transitions between AFCI and 2DTI/QSH (blue line) or AFCI and BI (green line) can be direct (as shown) or through intervening $C=0$ AF insulating phases (not shown).}
	\end{center}
\vskip-0.5cm
\end{figure}

A schematic phase diagram near the Weyl line
is shown in Fig.~1. There are at least three possible phases: 2DTI/QSH, trivial BI, and AFCI with $C=1$.
A continuous transition between 2DTI/QSH and BI is achieved by tuning the ${\cal T}$-symmetric $m_I$. Because of the vanishing density of states,
the Weyl points are stable against weak interactions. Thus, the line of Weyl points extends and terminates at a critical interaction strength beyond which they become gapped by $m_s\neq0$ due to the AF order along the spin direction of the Weyl fermions. The bifurcation of the Weyl line enables a new phase region, i.e. the AFCI. The transitions between AFCI and 2DTI/QSH or between AFCI and BI can be continuous, in which case one expects intervening $C=0$ AF insulating phases (not shown).
These transitions can also be discontinuous, going directly between 2DTI and BI or AFCI as depicted by the two first order lines in Fig.~1 that meet at the tricritical point.

Next, we study the proposed AFCI in a concrete model, namely the Kane-Mele Hubbard model,
\begin{eqnarray}
{H} &=& H_{\rm KM}+U\sum_{i} \hat{n}_{i,\uparrow}\hat{n}_{i,\downarrow},
\label{UV}
\end{eqnarray}
where $U$ is the onsite repulsion and $H_{KM}$ the well-known Kane-Mele Hamiltonian,
\begin{eqnarray}
{H_{\rm KM}} &=& -t \sum_{\langle i,j\rangle} c_{i}^{\dagger}c_{j}+i\lambda\sum_{\langle\langle ij\rangle\rangle} \nu_{ij}c_{i}^{\dagger}\sigma^zc_{j}
\nonumber \\
&+&\Delta\sum_{i}\xi_i c_{i}^{\dagger}c_{i}.
\label{km}
\end{eqnarray}
Here the electron operator is written in the spinor notation $c_{i}=(c_{i,\uparrow},c_{i,\downarrow})^T$. The term proportional to $\lambda$ is the SOI with $\nu_{ij}=(2/\sqrt{3})(\hat{\mathbf{d}}_1\times\hat{\mathbf{d}}_2)_z=\pm1$, where $\hat{\mathbf{d}}_1$ and $\hat{\mathbf{d}}_2$ are unit vectors along the two bonds that the electron traverses from site $j$ to its second nearest neighbor site $i$ on the honeycomb lattice as shown in Fig.~2(a). In Eq.~(\ref{km}), $\sigma^z$ is the Pauli matrix for the electron spin and $\Delta$ ($\xi_i=\pm1$) is a staggered sublattice potential. We set $t=1$ as the energy unit.
Consider first the electronic structure of $H_{KM}$. The ionic potential $\Delta$ breaks the inversion symmetry ${\cal I}$ that exchanges the two sublattices $A$ and $B$.
The degeneracy of $|k,\sigma\rangle$ and $|k,\bar{\sigma}\rangle$ is lifted, resulting in the split bands in Fig.~2(b)
except at the zone center $\Gamma$ point. The low energy states are determined by expanding momenta around $K=(\frac{4\pi}{3\sqrt{3}},0)$ and $K^\prime$ valley points. In the basis $(c_{kA\sigma}, c_{kB\sigma})$ with $\sigma=\pm$ for spins $\uparrow$ and $\downarrow$, we obtain
\begin{equation}
H_{\rm KM}(K,\sigma)=
\begin{bmatrix}
\Delta+3\sqrt{3}\sigma \lambda & k_x -i k_y \\
k_x +i k_y & -(\Delta+3\sqrt{3}\sigma\lambda)
\end{bmatrix}, \\
\end{equation}
and $H(K^\prime,\sigma)$ is given by the time-reversal of $H(K,\bar\sigma)$. When $\Delta$ and $\lambda$ have the same sign, the low energy physics is determined by $H(K,\downarrow)$ near $K$, and $H(K',\uparrow)$ near $K'$. Two gapless Weyl points at $K(\downarrow)$ and $K'(\uparrow)$ related by ${\cal T}$ emerge along the phase boundary $\Delta=3\sqrt{3}\lambda$ between the QSH and the trivial BI.

\begin{figure}
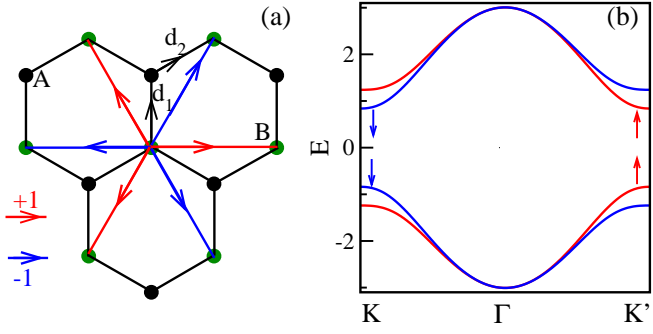

	\begin{center}
		\fig{3.4in}{band.eps}\caption{(a) Honeycomb lattice with second nearest neighbor SOI. $\nu_{ij}=+1$ for red lines and $-1$ for blue lines.(b) Spin up and spin down bands at $\lambda=0.2$ and $\Delta=0.2$.
}
	\end{center}
	\vskip-0.5cm
\end{figure}

At the mean-field level, the Hubbard interaction in $H$ is decoupled by Hartree and spin exchange self-energies
\begin{eqnarray}
H_{\rm MF}=H_{\rm KM}+\frac{U}{2}\sum_{i}(n_i \hat{n}_i-4\mathbf{S}_i\hat{\mathbf{S}}_i-\frac{n_i^2-4\mathbf{S}_i^2}{2})
\label{hmf}
\end{eqnarray}
where $\hat{n}_i=\sum_{\sigma}c_{i\sigma}^\dagger c_{i\sigma}$ and $\hat{\mathbf{S}}_i={1\over2}
\sum_{\alpha,\beta}c_{i\alpha}^\dagger\vec{\sigma}_{\alpha\beta}
c_{i\beta}$ are density and spin density operators with $ n_i $ and $\mathbf{S}_i$ their expectation values.
The second term in Eq.~(\ref{hmf}) renormalizes the ionic potential to
$
\Delta_{\rm eff}(U)=\Delta+\frac{1}{4}UQ,
$
where $Q=n_A- n_B$ measures the charge density wave (CDW) order.
Increasing $U$ reduces charge fluctuations and weakens the ${\cal I}$-breaking CDW order. The exchange interaction, the third term in Eq.~(\ref{hmf}), produces the AF order ${\bf m_s}=\frac{1}{2}({\bf S}_A-{\bf S}_B)$ for sufficiently large $U$.

\begin{figure}
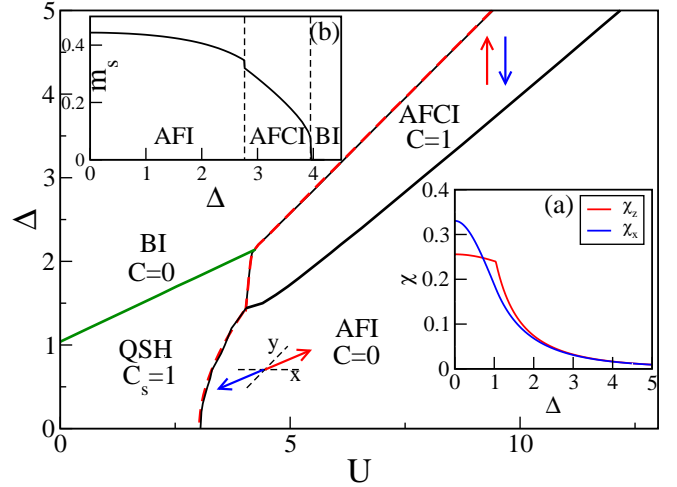

	\begin{center}
		\fig{3.4in}{phase0.eps}\caption{Phase diagram of the Kane-Mele Hubbard model at $\lambda=0.2$ showing the AFCI phase. The line of Weyl points (green solid line) marks the continuous transition between QSH and BI. All other meanfield phase boundaries are weakly first order (black solid lines). Magnetic phase boundaries coincide with those obtained by RPA (red dashed lines). A discontinuous spin-flop transition separates $z$-axis AFCI from $xy$-plane AFI. Inset (a): Bare spin susceptibilities as a function of sublattice potential $\Delta$ in the KM model. Inset (b): Magnetization $m_s$ versus $\Delta$ at $U=7.5$ with AFI-AFCI transition at $\Delta=2.77$ and AFCI-BI transition at $\Delta=3.96$.
}
	\end{center}
	\vskip-0.5cm
\end{figure}

The obtained phase diagram is shown in Fig.~3 on the $(U,\Delta)$-plane. 
There are four distinct phases: the QSH insulator with spin Chern number $C_s=\frac{1}{2}(C_\uparrow-C_\downarrow)=1$ and the BI already present in the noninteracting KM model; the $C=0$ $xy$-plane AF insulator (AFI) and the new $C=1$ $z$-axis AFCI.
Except for the continuous transition between QSH and BI,
all other transitions are weakly first order. The magnetic phase boundaries match well with those obtained using the conserving approximation and random phase approximation (RPA) \cite{baym1,baym2,flex} discussed in the supplemental material (SM).

In Fig.~3, the continuous boundary line of Weyl points between QSH and BI is determined by the gapless condition $\Delta_{\rm eff}(U)=3\sqrt{3}\lambda$. It terminates at the tricritical point $(\Delta_c,U_c)\simeq(2.14,4.25)$ beyond which the correlation-induced $C=1$ topological AFCI emerges with spin moments along the $z$-axis.
The phase structure in this region is consistent with the one proposed
in Fig.~1. The AFCI phase is bounded by weakly first order, direct transitions to QSH and CDW ordered BI.
These magnetic phase boundaries are approximately determined by $\Delta-\frac{1}{4}U\vert Q\vert\pm\frac{1}{2}U{\vert m_s\vert}=3\sqrt{3}\lambda$. The $C=1$ AFCI has several remarkable properties. (i) Being an interaction-driven topological insulator, it supports a single gapless chiral edge mode and $C=1$ QAHE as the result of spontaneous ${\cal T}$-symmetry breaking by AF long-range order. Due to the spin-valley locking and the topological mass for only one valley Weyl point, the gapless edge mode is spin-polarized despite bulk AF order. (ii) The AFCI occupies the part of the phase diagram with comparable $\Delta$ and $U/2$. In this regime, the charge fluctuations involving doubly-occupied sites (of energy cost $\sim U-2\Delta$) remain significant as $U$ drives the itinerant topological AF order; allowing the gapless edge state to carry both charge and spin of the quantized Hall current. Thus, the AFCI is fundamentally different from the AF Mott insulator in the large-$U$ and small-$\Delta$ regime, which is topologically trivial and described by spin-only low-energy theories such as the Heisenberg model.

The AFCI in Fig.~3 has spin moments aligned along the remaining spin-rotation invariant $z$-axis under the SOI. However, it is known that the AF Mott insulator at large-$U$ and $\Delta=0$ can be described by a Heisenberg model with an easy plane anisotropy \cite{hur,vaezi}. Fig.~3 shows that the phase at small $\Delta$ and large $U$ is indeed the $C=0$ $xy$-plane AFI, which is separated from the $z$-axis AFCI by a first order spin-flop transition obtained by comparing the energies of $xy$-plane and $z$-axis AF insulators. Note that because of the involvement of charge fluctuations, this is different from the usual spin-flop transitions in spin systems described by anisotropic Heisenberg models. To understand its origin, we calculate the $z$-axis and the $xy$-plane spin susceptibilities ($\chi_z$ and $\chi_x$) in the noninteracting KM model shown in the inset (a) of Fig.~3. The spin anisotropy clearly switches from easy-plane to easy-axis with increasing ionic potential $\Delta$. Consequently, the spin-flop transition can be described by the conserving approximation and RPA calculations (see SM) of the magnetic phase boundary between QSH and $xy$-plane AFI at small $\Delta$, QSH and $z$-axis AFCI at intermediate $\Delta$, and BI and $z$-axis AFCI at large $\Delta$, as shown in Fig.~3. Inset (b) in Fig.~3 shows the staggered magnetization as a function of $\Delta$ at $U=7.5$. The topological spin-flop transition demarcates the two distinct types of correlation-driven AF insulators.

The Hartree-Fock mean field theory tends to overestimate magnetic order. Thus, the phase diagram in Fig.~3 should be considered qualitative rather than quantitative. To verify that a strong-coupling treatment of $U$ does not alter the physical predictions, we use the spin-rotation invariant slave boson mean field theory \cite{kr86,li89,wolfle92,jiang14,newton}, equivalent to Gutzwiller approximation, to study the KM Hubbard model in Eq.~(\ref{UV}). Detailed discussions are given in the SM. In Fig.~4, the strong-coupling phase diagram is shown. Compared to the weak-coupling HF and RPA results shown in dashed lines, it is clear that, apart from the shift of magnetic phase boundaries to larger $U$, the phase structure remains unchanged.
The results confirm the existence of two kinds of AF insulators: the topologically trial AF Mott insulator and the topological AFCI with unquenched charge fluctuations.
\begin{figure}
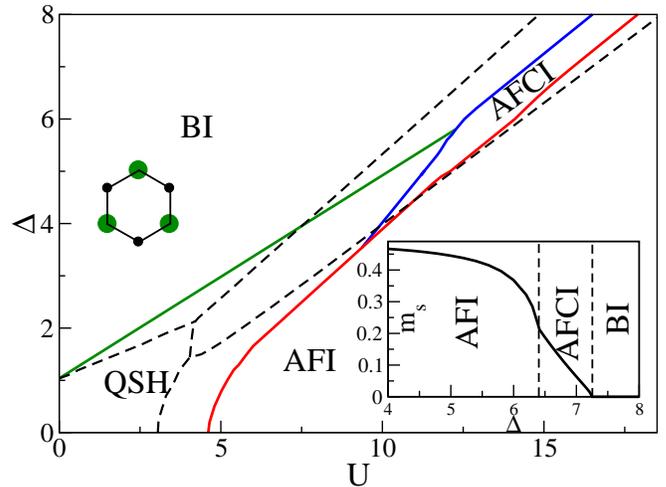

	\begin{center}
		\fig{3.4in}{gw.eps}\caption{Strong coupling phase diagram 
at $\lambda=0.2$. Green line: line of Weyl points. Blue line: phase boundary between QSH/BI and $C=1$ $z$-axis AFCI. Red line: phase boundary between QSH/AFCI and $xy$-plane AFI. Dashed lines are phase boundaries from Fig.~3. Inset: Magnetization $m_s$ as function of $\Delta$ at $U=15$.
}
	\end{center}
	\vskip-0.5cm
\end{figure}

We comment on the effects of Rashba SOI that can arise when the $z \rightarrow -z$ mirror symmetry is broken
on sample surfaces or
by coupling 2D systems to substrates. In the KM model, Rashba coupling $\lambda_R$ moves the phase boundary between QSH and BI to smaller $\Delta$ \cite{kane2}. We find that switching on Hubbard $U$ continues to produce the line of Weyl points and the transition to the $C=1$ AFCI. The phase diagram exhibits only small quantitative changes
as shown in the SM. However, since $S_z$ is no longer a conserved quantity, a small net ferromagnetic moment arises in the resulting ferrimagnetic states.

We have shown the existence of correlation-driven AFCI in non-centrosymmetric systems, where the QAHE emerges due to {\sl spontaneous} ${\cal T}$-breaking without net magnetization. This is in contrast to models of Chern insulators with ferromagnetism \cite{chang} or with explicit ${\cal T}$-breaking by external magnetic field \cite{huxiao} and skyrmion textures \cite{jelena}, or with complex hoppings \cite{liang}.
The truly ${\cal T}$-breaking, noncoplanar AF chiral spin density wave states \cite{taoli12,martin08,jiang15} can also produce AFCI with $C\ne0$.
The study of KM Hubbard model also shows that the topological AFCI can be obtained near the phase boundary between band and Mott insulators, when the inversion symmetry breaking potential is comparable to Hubbard $U$. Thus under proper ionic potential or crystal fields, the AFCI can be realized in systems with strong correlation and opens new directions for the search of QAHE in strongly correlated materials. Indeed, the underlying physics of the AFCI discussed here is general and model independent. In the SM, we show the emergence of AFCI and QAHE in a two-orbital Hubbard model \cite{wu07,wu08,wu14} realizable in cold atoms on optical lattices \cite{jordens,jotzu} where both the sublattice potential and SOI can be tuned optically \cite{optical,pan,demler}. Possible hosting 2D materials for the AFCI include the honeycomb lattice, noncentrosymmetric NiRuCl$_6$ with strong SOI \cite{zhou} and the deposition of heavy atoms with strong SOI on insulating magnetic substrates such as MnTe \cite{garrity} proposed recently by DFT calculations. Many transition metal dichalcogenides, such as
A(S,Se,Te)$_2$ where A=Mo,W
etc \cite{xiao,2d}, exist in stable monolayer ionic honeycomb structures without inversion symmetry \cite{eriksson}. Some of them, such as CrS$_2$, have been predicted to be AF \cite{eriksson} and 2DTI has been observed in others such as WTe$_2$ \cite{fei}. They are good candidate materials for such studies. There are also other candidate materials such as Na$_2$IrO$_3$, RuCl$_3$, LaNiO$_3$ \cite{kitaev1,kitaev2,xiao11,ying} that crystalize into hexagonal structures with the transition metal ions forming layered honeycomb structures along certain crystallographic directions.

\begin{acknowledgements}
This work is supported by the U.S. Department of Energy, Basic Energy Sciences Grant No. DE-FG02-99ER45747 (Z.W. and K.J.) and the Key Research Program of Frontier Sciences, CAS, Grant No. QYZDB-SSW-SYS012 (S.Z.). Z.W. thanks the hospitality of Aspen Center for Physics and the support of ACP NSF grant PHY-1066293.
\end{acknowledgements}

\newpage
\renewcommand{\theequation}{S\arabic{equation}}
\renewcommand{\thefigure}{S\arabic{figure}}
\renewcommand{\thetable}{S\arabic{table}}
\setcounter{equation}{0}
\setcounter{figure}{0}
\section*{ Supplementary Material }

\subsection{Conserving approximation and random phase approximation}

To determine the phase boundaries in weak coupling theories, we apply the conserving approximation, which was developed by Baym and Kadanoff for the electron gas \cite{sbaym1,sbaym2} and further extended to the fluctuation exchange approximation (FLEX) \cite{sflex}. We will discuss the procedure briefly, since it is well described by Bickers and Scalapino \cite{sflex}. The action can be written in the form of noninteracting and interaction parts as
\begin{eqnarray}
S&=&S_0+S_v \\
S_0&=&\beta\sum_{ij\sigma}c^{\dagger}_{i\sigma}(\frac{\partial}{\partial\tau}+H_0)_{ij}c_{j\sigma}\\
S_v&=&\frac{\beta}{2}\sum_{ij\sigma\sigma'}v^{\sigma\sigma'}_{ij}c^{\dagger}_{i\sigma}c^{\dagger}_{j\sigma'}c_{j\sigma'}c_{i\sigma}.
\end{eqnarray}
For the Hubbard model, the interaction $v^{\sigma\sigma'}_{ij}=U\delta_{ij}\delta_{\sigma'\bar{\sigma}}$. In a general self-consistent approximation, the two-particle interaction $v$ is replaced by a one-particle term describing the propagation in an external
field. This field is the irreducible one-particle self-energy $\Sigma_\sigma(ij)$, evaluated in some approximation such as the Hartree-Fock. Normally, the action can be divided into two parts describing the terms retained in the approximation and a residual. In the momentum-space representation,
\begin{eqnarray}
S_v&=&S_{sc}+\Delta S \\
S_{sc}&=&\beta\sum_{k\sigma}\Sigma_\sigma(k)c^{\dagger}_{k\sigma}c_{k\sigma}\\
\Delta S&=&S_v-S_{sc}
\end{eqnarray}
where $\beta=1/k_BT$ and the self-energy is a functional of the form $\Sigma=\Sigma(G;v)$ determined self-consistently through its dependence on the dressed Green's function $G$.
After self-consistently solving for $S_{sc}$, one could refine the approximation by treating $\Delta S$ perturbatively like random phase approximation (RPA). Such an expansion would employ self-consistently dressed $G$, rather than the bare $G_0$ used in RPA.

In the Kane-Mele Hubbard model, since $\Delta$ has already induced the charge density wave (CDW), we study the charge part self-consistently by Hartree approximation and then the spin sector using RPA by the $\Delta S$ expansion. The effect of the Hartree self-energy is to renormalize the ionic potential $\Delta$ to
\begin{equation}
\Delta_{\rm eff}(U)=\Delta+\frac{1}{4}UQ,
\end{equation}
where $Q=n_A-n_B$. Since there is only one interaction vertex $U$ in the spin sector, the spin susceptibility $\Pi_s$ for the three spin directions $s={x,y,z}$ can be obtained using the RPA formula,
\begin{eqnarray}
\Pi_s =\frac{\chi_s}{1-U\chi_s}
\end{eqnarray}
$\chi_s$ is spin susceptibility obtained using the Hartree-renormalized Green functions $G(\Delta_{\rm eff}(U))$. When $U=U_c^s=1/\chi_s$, $\Pi_s$ diverges, which corresponds to an instability toward a magnetic order of the spin $s$-component. Since system still has spin rotation symmetry in the $xy$ plane, we only need to compare $\chi_x$ and $\chi_z$ to determine the spin flop transition. We obtain
\begin{eqnarray}
\chi_{z}&=&\frac{1}{8}\sum_{k}\biggl[\frac{1}{\epsilon_{+}}+\frac{-(\Delta_{\rm eff}+\lambda\beta_k)^2+|\eta_k|^2}{\epsilon_{+}^3} \nonumber \\
&&+\frac{1}{\epsilon_{-}}+\frac{-(\Delta_{\rm eff}-\lambda\beta_k)^2+|\eta_k|^2}{\epsilon_{-}^3} \biggr], \label{chis} \\
\chi_{x}&=&\frac{1}{4}\sum_{k}\biggl [\frac{1}{\epsilon_{+}+\epsilon_{-}}+\frac{-\Delta_{\rm eff}^2+(\lambda\beta_k)^2+|\eta_k|^2}
{\epsilon_{+}^2\epsilon_{-}+\epsilon_{+}\epsilon_{-}^2}\biggr] \nonumber \\
\epsilon_{\pm}&=&\sqrt{(\Delta_{\rm eff} \pm \lambda \beta_k)^2+|\eta_k|^2},
\nonumber
\end{eqnarray}
where $\beta_k=2[\sin{k_x}-\sin{k_y}+\sin{(k_y-k_x)}]$ and $\eta_k=1+e^{-ik_x}+e^{-ik_y}$ with $k_{x,y}$ measured in units of reciprocal lattice vector $\vec b_{1,2}$ on the honeycomb lattice. The calculated spin susceptibilities $\chi_x$ and $\chi_z$ are plotted in the inset (a) of Fig.~3 in the main text as a function of $\Delta$ at $U=0$. For $U\neq0$, $\chi_x$ and $\chi_z$ depend on $U$ through $\Delta_{\rm eff}$ with the same functional dependence.

From the inset (a) of Fig.~3, we find that $\chi_x>\chi_z$ for $\Delta_{\rm eff}<0.71$, i.e. an easy plane anisotropy. In this region, the nonmagnetic QSH insulator becomes unstable against a magnetic transition to the $xy$-plane AFI with increasing $U$ as found in the phase diagram in Fig.~3. In the region where $\Delta_{\rm eff}>0.71$, $\chi_x<\chi_z$, which implies an easy $z$-axis anisotropy. The nonmagnetic QSH insulator and the BI show magnetic transitions to the $z$-axis AFCI as shown in Fig.~3. Note that the kink in $\chi_z$ in this region is due to the topological phase transition between the QSH insulator and BI at $\Delta_{\rm eff}=3\sqrt{3}\lambda$.
The phase boundaries are determined by the leading instabilities associated with the divergence of the RPA susceptibilities in Eq.~(S8), which are plotted in Fig.~3 in the main text using red dashed lines. They match precisely the corresponding magnetic phase boundaries obtained directly from the self-consistent Hartree-Fock mean-field theory.

\subsection{Slave Boson Mean Field Theory}

To treat the local Coulomb repulsion in a nonperturbative strong-coupling approach for SU(2) spins, we generalized the spin rotation invariant slave boson mean field theory \cite{kr86-s,li89-s,wolfle92-s,jiang14-s}, which is equivalent to Gutzwiller approximation, to include the presence of the SOI and obtained the phase diagram of the Kane-Mele Hubbard model. The starting point is to represent the local Hilbert space by a spin-1/2 fermion $f_\sigma$ and six bosons $e$, $d$, and $p_\mu$ ($\mu=0,1,2,3$) for the empty, doubly-occupied, and singly occupied sites respectively: $\vert0\rangle=e^\dagger \vert\text{vac}\rangle$,  $|\!\!\uparrow\downarrow\rangle=d^\dagger f_\downarrow^\dagger f_\uparrow^\dagger \vert\text{vac}\rangle$, and $\vert \sigma\rangle= {1\over\sqrt{2}} f_{\sigma^\prime}^\dagger p_\mu^\dagger \tau_{\sigma^\prime\sigma}^\mu \vert \text{vac}\rangle$ where ${\tau}^{1,2,3}$ and ${\tau}^0$ are Pauli and identity matrices.
The completeness of the Hilbert space and the equivalence between boson and fermion representations of the particle and spin density impose three local constraints:
\begin{eqnarray}
O_i&=&e_i^\dagger  e_i + p_{i0}^\dagger  p_{i0}  + \vec p_i^\dagger   \cdot \vec p_i  + d_i^\dagger  d_i -1=0, \nonumber \\
O_{i}^0&=&p_{i0}^\dagger  p_{i0}  + \vec p_i^\dagger   \cdot \vec p_i  + d_i^\dagger d_i - f_{i\sigma}^\dagger  f_{i\sigma}  = 0, \label{constraint1} \\
O_{i}^\alpha&=& p_{i0}^\dagger  p_{i\alpha}  + p_{i\alpha}^\dagger p_{i0} + i(\vec p_i^\dagger\times \vec p_i)_\alpha - f_{i\sigma}^\dagger\tau _{\sigma\sigma^\prime}^\alpha f_{i\sigma^\prime}  = 0,
\nonumber
\end{eqnarray}
where $\alpha=1,2,3$ corresponds to the $3$-directions of the electron spin. The Kane-Mele Hubbard Hamiltonian can thus be written as,
\begin{eqnarray}
H_{sb} &= & -t\sum_{\langle ij \rangle} \psi_{i}^\dagger  g_i^\dagger g_j \psi_j
+\sum_{\langle\langle ij \rangle\rangle} i\lambda \nu_{ij} \psi_{i}^\dagger  g_i^\dagger \sigma_z g_j \psi_j \nonumber \\
&+& \Delta\sum_{i}\xi_i \psi_{i}^{\dagger}\psi_{i}+U\sum_i {d_i^\dagger  d_i } \label{slavebosonh} \\
&-&\mu _0 \sum\limits_{i\sigma} f_{i\sigma }^\dagger  f_{i\sigma } + \sum_i\lambda _i O_i  + \sum_{i\mu} \lambda _{i\mu} O_{i}^\mu,
\nonumber
\end{eqnarray}
where the fermion spinor $\psi_i^\dagger=(f_{i\uparrow}^\dagger, f_{i\downarrow}^\dagger)$; $\lambda_i$ and $\lambda_{i\mu}$ ($\mu=0,1,2,3$) are Lagrange multipliers. The hopping renormalization factors $g_i$, $g_j$ are $2\times2$ matrices involving the boson operators \cite{li89-s,wolfle92-s,jiang14-s}.
\begin{equation}
g_i={\bf L}_i^{-1/2}(e_i^\dagger {\bf p}_i+{\bf {\overline p}}_i^\dagger d_i){\bf R}_i^{-1/2},
\label{zfactor}
\end{equation}
where ${\bf p}$ is a $2\times2$ matrix with $p_{\sigma\sigma^\prime}^\dagger={1\over\sqrt{2}}\sum_{\mu} p_\mu^\dagger \tau_{\sigma\sigma^\prime}^\mu$, ${\bf L_i}=(1-d_i^\dagger d_i)\tau_0-{\bf p_i}^\dagger {\bf p}_i$, ${\bf R}_i=(1-e_i^\dagger e_i)\tau_0-{\bf {\overline p_i}}^\dagger {\bf{\overline p}}_i$ with ${\bf{\overline p}}_i={\hat{\bf T}}{\bf p}_i{\hat{\bf T}}^{-1}$ the time-reversal transformed ${\bf p}_i$. The saddle-point solution corresponds to condensing all boson fields $(e_i,d_i,p_{i\mu},\lambda_i,\lambda_{i\mu})$ and determining their values self-consistently by minimizing the ground state energy. The latter gives rise to the following self-consistency equations at each site,
\begin{eqnarray}
\frac{{\partial T }}{{\partial e_\ell }} &+& 2\lambda_\ell e_\ell  = 0,
\nonumber \\
\frac{{\partial T }}{{\partial d_\ell }} &+& (2\lambda_\ell  - 4\lambda_{\ell0} + 2U)d_\ell  = 0,
\\
\frac{{\partial T }}{{\partial p_{\ell0} }} &+& 2\lambda_\ell p_{\ell0} - 2\sum_\mu\lambda_{\ell\mu} p_{\ell\mu} = 0,
\nonumber \\
\frac{{\partial T }}{{\partial p_{\ell\alpha}}}
&+&2\lambda_\ell p_{\ell\alpha} - 2\lambda_{\ell0} p_{\ell\alpha}-2\lambda_{\ell\alpha} p_{\ell0}=0,
\nonumber
\end{eqnarray}
where $T$ is the quantum averaged kinetic energy,
\begin{equation}
T=-t\sum_{\langle ij \rangle} \langle \psi_{i}^\dagger g_i^\dagger g_j \psi_j\rangle
+\sum_{\langle\langle ij \rangle\rangle} i\lambda \nu_{ij} \langle \psi_{i}^\dagger  g_i^\dagger \sigma_z g_j \psi_j \rangle. \nonumber
\end{equation}
These equations, together with the quantum averaged constraints in Eq.~(\ref{constraint1}), are solved numerically
by discretizing the reduced zone with typically $600\times600$ points to allow accurate determinations of the ground state properties.
To achieve better convergence, we solve the self-consistent equations using Newton's method discussed in detail in Ref.~\cite{snewton}. The obtained strong coupling phase diagram is shown in Fig.~4 in the main text.

\begin{figure}
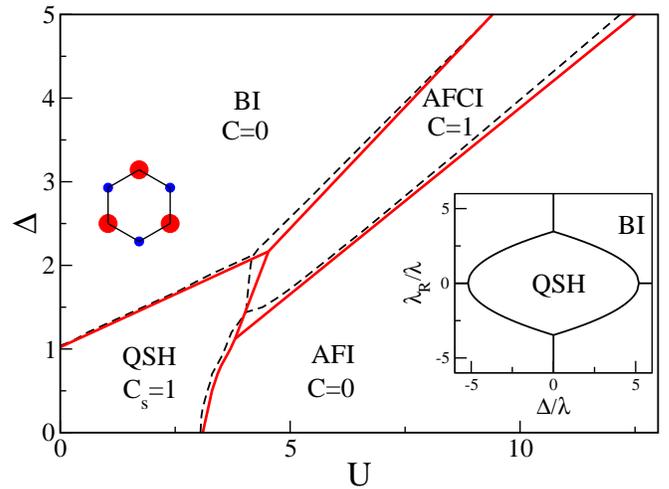

	\begin{center}
		\fig{3.4in}{phase_rc.eps}\caption{Phase diagram of the Kane-Mele Hubbard model including both the SOI $\lambda=0.2$ and the Rashba coupling $\lambda_R=0.1$. Dashed lines are phase boundaries in the absence of Rashba coupling. Inset: the phase diagram of non-interacting Kane-Mele model showing the evolution of phase boundaries with $\lambda_R$ between the QSH and BI.}
	\end{center}
	\vskip-0.5cm
	\label{figS1}
\end{figure}

\subsection{Effects of Rashba Coupling}	

When the $z \rightarrow -z$ mirror symmetry is broken by a perpendicular electric field on the sample surface or by coupling the 2D system to a substrate, the Rashba spin orbit coupling term can arise. In the Kane-Mele model on the honeycomb lattice, it is given by
\begin{eqnarray}
H_R=i\lambda_R \sum_{\langle ij\rangle} c_{i}^{\dagger}(\vec{\sigma} \times \mathbf{\hat{d}}_{ij})_zc_{j}.
\end{eqnarray}
Adding the Rashba term completely breaks spin-rotation symmetry, such that $S_z$ is no longer a conserved quantity. In the noninteracting limit $U=0$, the phase diagram in the plane spanned by $\Delta/\lambda$ and $\lambda_R/\lambda$ was obtained by Kane and Mele \cite{kane2-s}, which is shown in the inset of Fig.~S1. The phase boundary between the QSH and BI moves to smaller $\Delta$ with increasing $\lambda_R$.
We calculated the phase diagram of the Kane-Mele Hubbard model in the presence of Rashba coupling within the Hartree-Fock mean-field theory, which is shown in Fig.~S1 at $\lambda=0.2$ and $\lambda_R/\lambda=0.5$.
Comparing to the dashed phase boundary lines at $\lambda_R=0$, we conclude that the phase structure remains the same with only small quantitative shifts of the phase boundaries. Since $S_z$ is no longer conserved, the AF ordered states contain small net ferromagnetic moments and are therefore weakly ferrimagnetic.

\subsection{AFCI in the two-orbital Hubbard model on the honeycomb lattice}

The physics associated with the emergence of the AFCI discussed here is quite general and can be realized in other models besides the Kane-Mele Hubbard model. As an important example, we discuss the $(p_x,p_y)$ two-orbital Hubbard model on the honeycomb lattice, since it can be realized and has been studied in connection to the ultra-cold atoms on optical lattices \cite{swu07,swu08,swu14}.
The optical potential around the minima at the lattice points is locally harmonic and can be used to produce a large band gap that well separates the bands associated iwth the $s$ and $p$ orbitals. By imposing strong laser beams along the $z$ direction, the band of the $p_z$ orbital can be pushed to very high energies. As a consequence, an ideal $(p_x,p_y)$ two-orbital system is realized on the artificial honeycomb optical lattice. Furthermore, the two $e_g$ orbitals of the transition metal $d$-electrons can be realized on the honeycomb lattice in, for example, the bilayer LaNiO$_3$ along the (111) surfaces \cite{sying}, which behave in a similar fashion to the $(p_x,p_y)$ two-orbital model.

The Hamiltonian of the two-orbital model can be written as $H=H_0+H_I$ where $H_0$ is the noninteracting part and $H_I$ describes the local Coulomb interactions. The noninteracting $H_0=H_t+H_{SOC}+H_{V}$ where $H_t$ is the tight-binding part describing the nearest neighbor hopping of the electrons in the $(p_x,p_y)$ orbitals on the honeycomb lattice; $H_{SOC}$ accounts for the SOC $\lambda$,
\begin{equation}
H_{SOC}= -\lambda \sum\limits_{i}(i p_{i\uparrow x}^\dagger p_{i\uparrow y}-ip_{i\downarrow x}^\dagger p_{i\downarrow y} +h.c);
\end{equation}
and $H_V$ for the tunable ionic potential $V$ on the optical lattice,
\begin{equation}
H_{V}= V \sum\limits_{i}({\hat{n}_{i,A}-\hat{n}_{i,B}}),
\end{equation}
where $\hat{n}_{i,A/B}=\sum_{\sigma\alpha}p_{i\sigma\alpha,A/B}^\dagger p_{i\sigma\alpha,A/B}$ is the electron density operators for both spins $\sigma=\uparrow,\downarrow$ and orbitals $\alpha=x,y$ on the $A/B$ sublattices. To describe the hopping part,
we introduce the four-component, orbital-sublattice spinor representation in momentum space defined as \cite{swu14},
\begin{equation}
p_{k\sigma}=[p_{\sigma x,A}(k),p_{\sigma y,A} (k),p_{\sigma x,B}(k),p_{\sigma y,B}(k)]^T,
\end{equation}
and write $H_t=\sum_{k\sigma}p_{k\sigma}^\dagger H_{t\sigma}(k)p_{k\sigma}$, where
\begin{equation}
H_{t\sigma}(k)=
\begin{bmatrix}
0 & T \\
T^+ & 0
\end{bmatrix},
\end{equation}
and
\[
T=
\begin{bmatrix}
t_\pi+\frac{3t_\sigma+t_\pi}{4}(e^{ik_x}+e^{ik_y}) & \frac{\sqrt{3}(t_\sigma-t_\pi)}{4}(e^{ik_x}-e^{ik_y})
\\ \frac{\sqrt{3}(t_\sigma-t_\pi)}{4}(e^{ik_x}-e^{ik_y}) & t_\sigma+\frac{t_\sigma+3t_\pi}{4}(e^{ik_x}+e^{ik_y})
\end{bmatrix}.
\]
Here the momenta $k_x$ and $k_y$ are measured along the reciprocal lattice vectors $\vec{b}_{1/2}$ of the honeycomb lattice, and $t_{\sigma}$ and $t_{\pi}$ are the bonding strengths (hopping integrals) of the $\sigma$ and $\pi$ orbitals. Clearly, the $S_z$ component of the electron spin is conserved in the noninteracting Hamiltonian $H_0$. As in the study of the $(p_x,p_y)$ two-orbital model on the optical lattice \cite{swu07,swu08,swu14}, we ignore the hopping of the $\pi$-bonding orbital and set $t_\sigma=1$ as the energy unit.
More detailed discussions of the model can be found in Refs.~\cite{swu07,swu08,swu14}.

For the interacting part $H_I$, we consider the standard two-orbital Hubbard interactions
\begin{align}
	H_I&=U\sum_{i,\alpha}\hat{n}_{i\alpha\uparrow}\hat{n}_{i\alpha\downarrow}
	+\left(U'-{1\over 2}J\right)\sum_{i,\alpha<\beta}\hat{n}_{i\alpha}\hat{n}_{i\beta}
	\label{hi} \\
	&-J_H\sum_{i,\alpha\neq\beta}{\bf S}_{i\alpha}\cdot {\bf S}_{i\beta}
	+J_H\sum_{i,\alpha\neq\beta}p^\dagger_{i\uparrow\alpha}
	p^\dagger_{i\downarrow\alpha}p_{i\downarrow\beta}p_{i\uparrow\beta},
	\nonumber
\end{align}
where the intraorbtial $U$ and interorbital $U^\prime$ Coulomb repulsions are related by the Hund's coupling $J_H$ according to $U=U'+2J_H$.
In the presence of SOC, the Hartree and exchange self energies introduced by $H_I$ depend on the full spin-orbital dependent density operator,
\begin{equation}
\hat n^{\alpha\beta}_{i\sigma\sigma'}= p^\dagger_{i\sigma\alpha} p_{i\sigma'\beta}.
\end{equation}
Local physical quantities such as the orbital-dependent density and spin density operators can be expressed as
\begin{equation}
\hat{n}_{i,\alpha}=\sum\limits_{\sigma} \hat{n}_{i,\sigma \sigma}^{\alpha \alpha},    \quad \hat{m}_{i,\alpha}^\mu=\sum\limits_{\sigma \sigma'} s_{\sigma \sigma'}^\mu\hat{n}_{i,\sigma \sigma^\prime}^{\alpha \alpha},
\end{equation}
where $s^\mu$, $\mu=x,y,z$, are the Pauli matrices. The other exchange self-energies include the spin-conserved and spin-flip orbital off-diagonal $L_{i,\alpha\beta}'$ and $L_{i,\alpha\beta}''$, as well as the spin-conserved and the spin-flip spin-orbital $R_{i,\alpha\beta}'$ and $R_{i,\alpha\beta}''$ contributions,
\begin{eqnarray}
\hat{L'}_{i,\alpha \beta}&=&\sum\limits_{\sigma} \hat{n}_{i,\sigma \sigma}^{\alpha \beta}, \quad \hat{L''}_{i,\alpha \beta}=\sum\limits_{\sigma} \hat{n}_{i,\sigma \bar{\sigma}}^{\alpha \beta}, \nonumber  \\
\hat{R'}_{i,\alpha \beta}&=&\sum\limits_{\sigma} \sigma \hat{n}_{i,\sigma \sigma}^{\alpha \beta}, \quad \hat{R''}_{i,\alpha \beta}=\sum\limits_{\sigma} \sigma \hat{n}_{i,\sigma \bar{\sigma}}^{\alpha \beta},
\nonumber
\end{eqnarray}
where $\alpha\ne\beta$. Including all Hartree and exchange self-energies in a fully self-consistent meanfield calculation, we obtain
the phase diagram as functions of the Hubbard $U$ and the ionic potential $V$ shown Fig.~S2 for SOC $\lambda=0.4$ and Hund's rule coupling $J_H=0.1U$. Remarkably, the phase diagram is very similar to the one obtained for the Kane-Mele Hubbard shown in Fig.~3 in the main text. Specifically, the phase boundary, i.e. the line of 2D Weyl points separating the 2DTI/QSH 
and the trivial band insulator, extends to a critical value of $U$ and bifurcates to give rise to the $C=1$ AFCI with ordered moments along the $z$-axis. At large-$U$ and a fixed $V$, the AFCI undergoes a topological spin-flop transition to the topologically trivial AF Mott insulator with ordered moments in the $xy$-plane.

These results further support that the underlying physics for the emergence of the topological AFCI discussed in main text is rather general and not model dependent. Moreover, its existence in the $(p_x,p_y)$ two-orbital Hubbard model encourages the search for the AFCI in ultracold atoms on optical lattices.


\begin{figure}
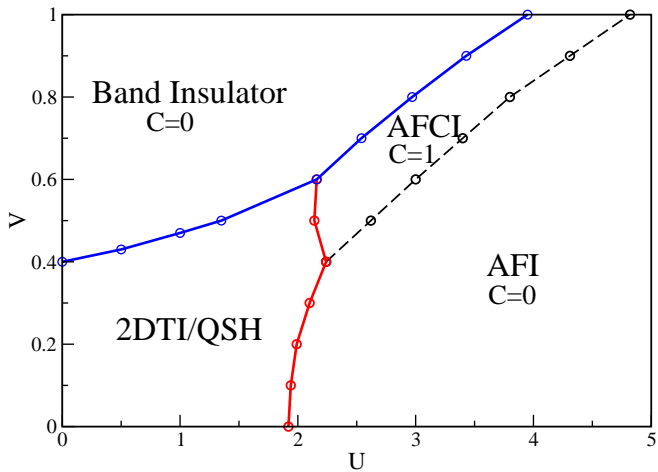

	\begin{center}
		\fig{3.4in}{lamb.eps}
		\caption{Phase diagram of the $(p_x,p_y)$ two-orbital Hubbard model spanned by the Hubbard $U$ and the sublattice potential $V$ obtained for the SOC $\lambda=0.4$ and Hund's rule coupling $J_H=0.1U$. There are four phases: the 2DTI/QSH, the topologically trivial band insulator, the AFCI with Chern number $C=1$, and the AF insulator (AFI) with $C=0$.}
	\end{center}
	\vskip-0.5cm
\end{figure}

\end{document}